# Epitaxial growth and magnetic characterization of EuSe thin films with various crystalline orientations

Ying Wang[1*], Xinyu Liu,[2*] Seul-Ki Bac,[2] Jacek K. Furdyna,[2] Badih A. Assaf,[2] Maksym Zhukovskyi,[3] Tatyana Orlova,[3] Neil R Dilley[4], Leonid P. Rokhinson[1,4,5]

[1] Department of Physics and Astronomy, Purdue University, West Lafayette, IN 47907

[2] Department of Physics, University of Notre Dame, Notre Dame IN 46556

[3] Notre Dame Integrated Imaging Facility, University of Notre Dame, Notre Dame, IN 46556, USA

[4] Birck Nanotechnology Center, Purdue University, West Lafayette, IN 47907

[5] Department of Electrical and Computer Engineering, Purdue University, West Lafayette, IN 47907

**Abstract:** We report different growth modes and corresponding magnetic properties of thin EuSe films grown by molecular beam epitaxy on $BaF_2$, $Pb_{1-x}Eu_xSe$, GaAs and $Bi_2Se_3$ substrates. We show that EuSe growth predominantly in (001) orientation on GaAs(111) and $Bi_2Se_3$, but along (111) crystallographic direction on $BaF_2$ (111) and $Pb_{1-x}Eu_xSe$ (111). High resolution transmission electron microscopy measurements reveal an abrupt and highly crystalline interface for both (001) and (111) EuSe films. In agreement with previous studies, ordered magnetic phases include antiferromagnetic, ferrimagnetic and ferromagnetic phases. In contrast to previous studies, we found strong hysteresis for the antiferromagnetic-ferrimagnetic transition. An ability to grow epitaxial films of EuSe on $Bi_2Se_3$ and of $Bi_2Se_3$ on EuSe enables further investigation of interfacial exchange interactions between various phases of an insulating metamagnetic material and a topological insulator.

## Introduction

Eu chalcogenides are magnetic insulators that host one of the highest magnetic saturation moments per atom, owing to a partially filled Eu 4f-shell. A renewed interest in these materials is motivated by a possibility to induce strong Zeeman spin-splitting when interfaced with topological insulators and superconductors. Topological insulators (TIs) are characterized by the presence of topologically protected gapless surface states (TSSs) with spin-momentum locking [1] that are both a blessing and a curse for TI-based devices: on one hand, they are protected by time-reversal symmetry; on the other, the geometry of TSS cannot be constrained by simple lithographic techniques. There are many applications where selective gapping of TSS would be desired, for example the quantum anomalous Hall effect requires gapping of top and bottom surfaces [2-3], spintronic devices may benefit from an ability to gap the side TSS and, thus, electrically disconnect top and bottom TSSs [4], while in TI-based topological superconductors selective gaping of TSS is required in order to localize non-Abelian excitations [5].

Globally, time-reversal protection of TSS can be lifted by strong external magnetic fields [6] or via bulk doping of a TI with magnetic impurities [7-8]. It has been suggested that local control of TSS can be achieved by coupling a TI to a magnetic material [9] [10]. Opening of an exchange gap in the Dirac spectrum of TSS has been reported in $Bi_2Se_3$/EuS heterostructures, where ferromagnetic insulator EuS proximity-induced ferromagnetism in the TI surface[11] . Polarized neutron reflectometry detected interfacial magnetization in these heterostructures up to room temperature, 20 times higher than the Curie temperature $T_c = 17$ K of the bulk EuS [12]. Such large $T_c$ enhancement has not been seen in subsequent studies [13]. Gapping of TSS by exchange-coupling to an antiferromagnetic insulator has an advantage of reduced stray magnetic fields from the bulk of the magnetic material and has been demonstrated in magnetically doped TI interfacing $Cr_2O_3$ [14] and CrSb [15]. There are currently no reports of both top and bottom TSS modulations in non-magnetic TIs, which motivates further investigation of TI/magnetic material interface formation and the nature of interfacial magnetic exchange.

EuSe is an insulating metamagnetic material with almost perfect cancellation of the nearest and next-nearest neighbor interactions between localized magnetic moments on the half-filled 4f levels of $Eu^{2+}$ ions [16]. This cancelation leads to a phase diagram which includes several antiferromagnetic (AF-I ↑↓↑↓, AF-II ↑↑↓↓), ferrimagnetic (FiM ↑↓↑) and ferromagnetic (FM) phases in the bulk, where magnetization vectors are confined to the (111) plane of the NaCl-type crystal structure (as shown in Fig. 1). It would thus be of great interest to control orientation of EuSe in MBE growth, to enable proximity with ferromagnetic (111) and antiferromagnetic (100) surfaces. The relative strength of magnetic phases and positions of phase boundaries are also highly sensitive to strain and can be tuned during epitaxial growth [17] [18]. The magnetic properties of bulk EuSe and thin layers grown on IV-VI materials have been studied in the past [17] [19] [20].

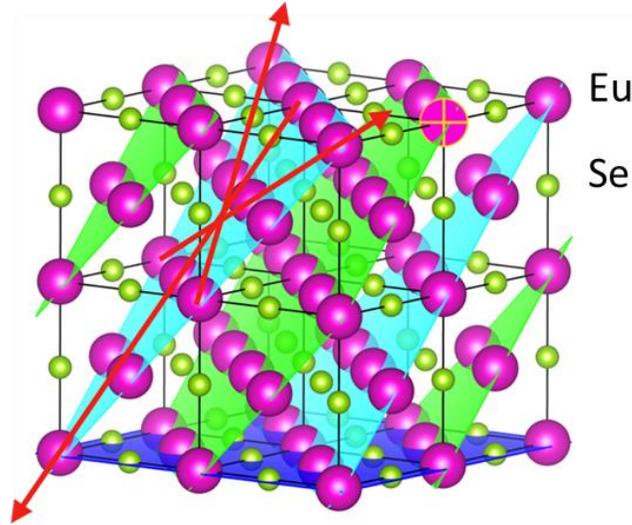

Fig. 1. Crystal structure of EuSe. Magnetization vector is laying within one of the four (111) planes (T-domains), red arrows indicate three possible in-plane magnetizations for each T-domain (S-domains).

In this work we report growth and magnetic characterization of thin EuSe films grown on various substrates including TIs. We synthesize EuSe on lattice-mismatched GaAs(111)B and $Bi_2Se_3$, and compare films properties to the ones grown on a lattice-matched $BaF_2$(111) and slightly lattice-mismatched $Pb_{1-x}Eu_xSe$ (111). We show that on GaAs(111)B and $Bi_2Se_3$ EuSe grows predominantly in (001) direction. EuSe films grown on $BaF_2$(111) and $Pb_{1-x}Eu_xSe$ (111) grow along (111) crystallographic direction consistent with previous studies. We performed SQUID magnetometry and constructed magnetic phase diagrams for (001) and (111) EuSe films.

## Material growth

The EuSe thin epilayers were grown by molecular beam epitaxy (MBE) on three types of substrates and buffer layers: (SA) lattice mismatched GaAs, (SB) nearly-lattice matched $BaF_2$ (111), (SC) $Pb_{1-x}Eu_xSe/BaF_2$ (111) pseudo-substrates, and (SD) 2-dimensional van-der-Waals $Bi_2Se_3$ epilayers grown on sapphire. Table 1 summarizes parameters of the devices. High-purity PbSe compound flux, and Eu, Bi, and Se elemental fluxes are obtained from standard effusion cells. The growth is monitored by *in situ* reflection high energy electron diffraction (RHEED). During the growth process, the substrate temperature is kept at 300°C and the growth is carried out under Se-rich condition. The Se flux is kept constant at the beam equivalent pressure (BEP) ratio of Se:Eu $\sim 4 - 20$. Under these conditions the growth rate of EuSe ($\sim 1 - 2$ nm/mins) is controlled by the Eu flux. Various substrate treatments are carried out prior to the deposition of EuSe. The $BaF_2$ and sapphire substrate are annealed *in-situ* at 650-700°C for at least 1 hour for thermal cleaning.

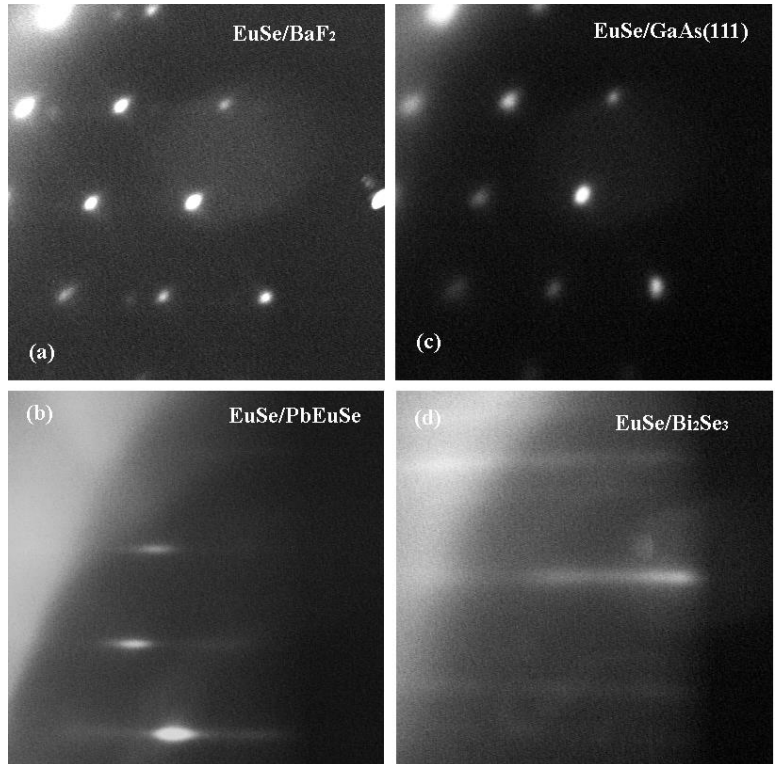

Fig. 2. RHEED patterns of EuSe films grown on different substrates: (a) BaF$_2$ (111), (b) Pb$_{1-x}$Eu$_x$Se/BaF$_2$ (111), (c) GaAs (111)B, and (d) Bi$_2$Se$_3$.

We found that direct growth of EuSe on BaF$_2$ (SB) is quasi-two dimensional and yields a blurred and spotty RHEED pattern, as seen in Fig. 2(a), and a rough surface. An introduction of a Pb$_{1-x}$Eu$_x$Se ($x > 0.42$) buffer layer restores epitaxial 2D growth (SC), as indicated by the streaky RHEED pattern in Fig. 2(b). Two types of Pb$_{1-x}$Eu$_x$Se buffers are used in this work: a multi-layer PbEuSe graded buffers and PbSe/EuSe short-period superlattice (SL). It is found that the SL results in improved surface roughness compared to a graded buffer (0.77 nm vs 0.98 nm rms). EuSe film grown on PbEuSe graded buffers is compressively strained.

For GaAs (111)B substrates (SA) a specific Se treatment was carried out prior to the growth. First, a GaAs (111)B substrate is heated to ~570°C to remove native oxide. Next, the substrate is annealed under Se flux (~$2 \times 10^{-6}$ torr) at 600°C for 20 minutes to obtain streaky (1×1) RHEED pattern. Se passivation smoothes the surface and terminates it with Se bonds. EuSe grown on GaAs (111)B substrate has two initiation processes. The direct growth often yields a 3-dimentional growth mode with a rough surface, as seen in Fig. 2(c). Alternatively, a specific atomic-layer-epitaxy (ALE) growth process carried out to initiate the growth achieves a layer-by-layer growth mode: the growth is initiated by a depositions of 6 periods of Se monolayers followed by a monolayer of Eu at lower temperature (200°C). The substrate is then gradually heated to 300°C, and a nice streaky RHEED pattern (not shown, similar to Fig. 2(d)) appears prior to the MBE growth of EuSe film. The latter process with higher Se:Eu flux ratio yields a uniform EuSe in (001) growth direction with a 12-fold symmetry RHEED pattern, suggesting that four EuSe {100} planes are parallel (aligned) to three {110} plane of GaAs (111) surface similar to what has been recently reported for the growth of PbSe of GaAs [21] (see also Fig. S2(b) in the Supplemental Material).

| Sample | growth sequence | thickness (nm) | $R_a$(nm) | growth | $T_N$(K) | Strain |
|---|---|---|---|---|---|---|
| SA1 | GaAs(111)B/EuSe | 120 | 2.96 | (001) | 5.1 | -0.001% |
| SA2 | GaAs(111)B/EuSe | 120 |  | Mixed (001) & (111) | 5.4 | +0.09% |
| SB | BaF$_2$(111)/EuSe | 90 | 2.58 | (111) | 4.9 | -0.09% |
| SC | BaF$_2$(111)/ PbSe/ Pb$_{0.8}$Eu$_{0.2}$Se/ Pb$_{0.575}$Eu$_{0.425}$Se/ EuSe/Bi$_2$Se$_3$ | 64/57/43 /20/28 | 0.979 | (111) | 5.5 | +0.65% |
| SD1 | Sapphire/Bi$_2$Se$_3$/EuSe | 16.6/20 | 0.886 | (001) | 4.4 | -0.08% |
| SD2 | GaAs(111)B/Bi$_2$Se$_3$/EuSe | 18.8/8.2 | 1.74 | (001) with (111) inclusions | 4.4 | +0.11% |
| SF | GaAs (111)B/ EuSe/Bi$_2$Se$_3$/EuSe | 9.9/16.5/ 6.7 | 1.96 | bottom: (001) & (111), top: (001) |  | +0.05% |

Table I. Properties of EuSe films studied in this work. Films thickness is determined from MBE growth calibration and confirmed for several films by TEM imaging. RMS surface roughness Ra is extracted from atomic force microscopy (AFM) images. The dominant crystal orientation is determined by XRD. Occasional (111) inclusions are detected by TEM but not by XRD. The Néel temperature is extracted from temperature dependence of magnetic susceptibility. The out-of-plane strain is calculated based on the out-of-plane lattice constant measured by XRD.

We also performed growth of EuSe on Bi$_2$Se$_3$, itself grown on sapphire *c*-plane (sample SD1) or GaAs (111)B (SD2) substrates. 15-20 nm thick Bi$_2$Se$_3$ were grown at 300°C under Se rich condition as in ref.[22] [23] [24], followed by the EuSe growth. Under Se-rich condition (Se:Eu > 10), the RHEED pattern is blurry but streaky, suggesting a layer by layer growth mode (see Fig. 2(d)). The process yields EuSe in (001) growth direction with a pseudo 24-fold symmetry rotations, confirmed by the *φ*-scan of {224} planes of (001) EuSe (see Supplementary materials). Lastly, a tri-layer

EuSe/Bi$_2$Se$_3$/EuSe (SF) is grown on GaAs(111) following the combined approach used for SA1 to grow EuSe on GaAs and SD2 to grow Bi$_2$Se$_3$ and EuSe.

## Structural characterization

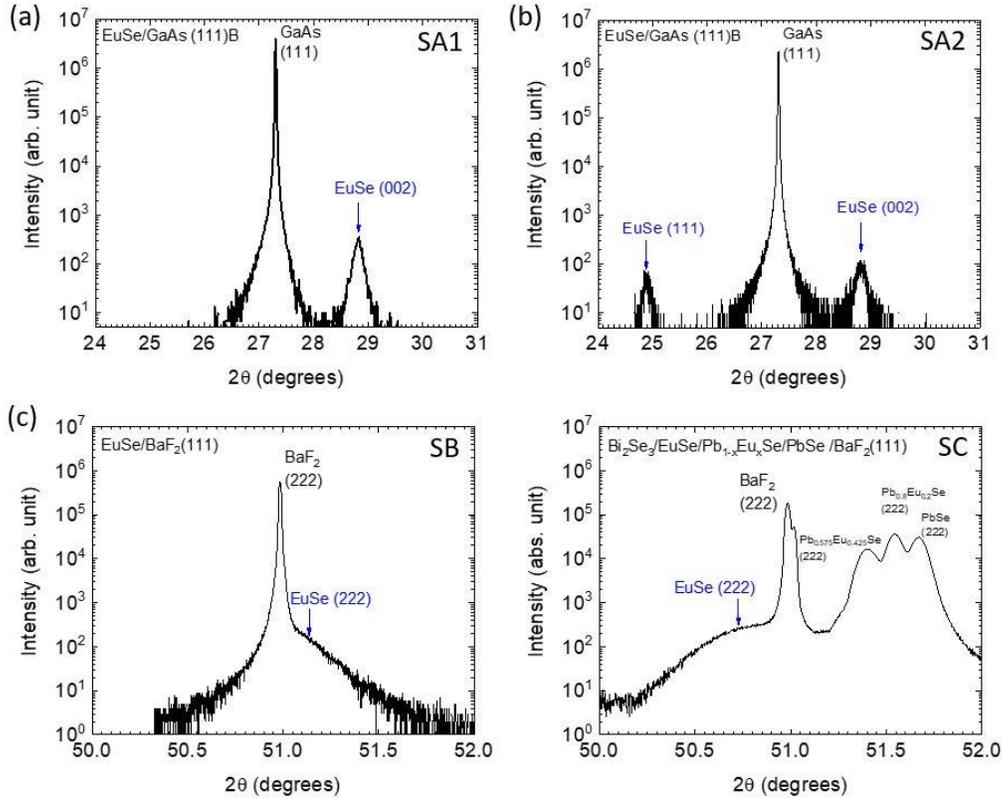

Fig. 3. Structural characterization of EuSe growth on different substrates. (a)-(d) HRXRD of EuSe films grown on (a) (b) GaAs (111)B, (c) BaF$_2$ (111), (d) Pb$_{1-x}$Eu$_x$Se/BaF$_2$ (111) substrates.

Structurally, epi-layers are characterized by high resolution X-ray diffraction (HRXRD) using 1.5406 Å $Cu\_K_{\alpha 1}$ radiations and transmission electron microscopy (TEM). Figs. 3 and 4 show X-ray diffraction characterizations revealing growth direction and strain of EuSe thin films grown on different materials. The growth of EuSe on GaAs (111)B (SA) has two modes: a mixed phase with both EuSe (111) and EuSe (001); and a single phase with EuSe (001) shown in Fig. 3 (a) (b) for two different samples. The growth direction depends on the Se:Eu ratio: for Se:Eu $>$ 10 a (001) growth is preferred (SA1), while for Se:Eu$\sim$4, a (001) growth with (111) inclusions is observed (SA2), as can be seen from both XRD spectrum and TEM images. The structural properties of EuSe grown on GaAs (111)B and the corresponding interface alignments are discussed in Supplementary materials (Figs. S2 and S3).

The XRD pattern of SB grown on BaF$_2$ (111) is shown in Fig. 3(c). A (111) oriented EuSe layer that is negligibly strained (<0.09%) is obtained in this case. The XRD pattern from SC grown on a Pb$_{1-}$

$_x$Eu$_x$Se buffer layer is shown in Fig. 3(d). This layer is also (111) oriented but slightly compressively strained in the plane and tensile strained in the (111) direction (∼ −0.6%). Fig. 4(a) shows the XRD patterns of EuSe grown on Bi$_2$Se$_3$ (SD1, SD2 and SF). A strong (002) EuSe peak is observed in all samples indicating that all EuSe (including both the top and the bottom layer in SF) are predominantly (001) oriented.

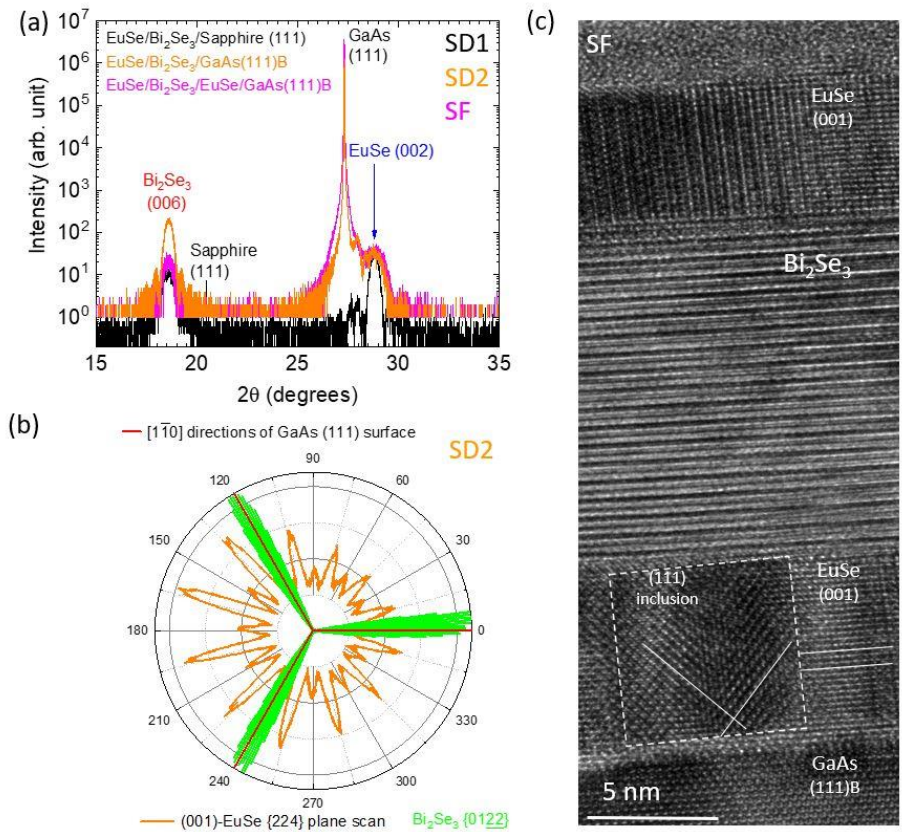

Fig. 4. Structural characterization of EuSe growth on Bi$_2$Se$_3$. (a)-(c) (a) High-resolution XRD ω-2θ scan of the EuSe film grown on Bi$_2$Se$_3$/GaAs(111)B see (002) and (004) Bragg peaks of EuSe, and a series of Bi$_2$Se$_3$ peaks. (b) φ-scan of the (224) EuSe Bragg peak with (001) growth direction, and φ-scan of the {01$\underline{22}$} Bi$_2$Se$_3$ Bragg peak (c) High-resolution TEM image for the EuSe/Bi$_2$Se$_3$/EuSe epilayers grown on GaAs (111)B substrate shows a predominantly [001] oriented growth of EuSe on GaAs (111)B with some (111) inclusions.

The lattice constant extracted from these peaks (6.186 Å) suggests EuSe is grown along the (001) direction and is fully relaxed on the c-plane of Bi$_2$Se$_3$ (001). The Bi$_2$Se$_3$ (006) peak also appears in the data and yield a lattice constant (28.511 Å) close to that of bulk Bi$_2$Se$_3$. The XRD pattern of the tri-layer is also shown in Fig. 4(a) and yields Bragg peaks that match exactly with those observed in bi-layers. In this tri-layer sample both Bi$_2$Se$_3$ surfaces are in proximity with (001) surfaces of EuSe.

A φ-scan of {224} planes of (001) EuSe from SD2 exhibits a pseudo 24-fold symmetry shown in Fig. 4 (b): four EuSe {100} planes are dominantly parallel (aligned) to three {110} planes of Bi$_2$Se$_3$

(001) surface, while four EuSe {110} planes are parallel (aligned) to three {110} planes of Bi$_2$Se$_3$ (001) surface. This is compared with the well-known three-fold symmetric pattern obtained from Bi$_2$Se$_3$. Thus, while the out-of-plane orientation of EuSe is predominantly along the (001) direction, two types of in-plane alignments are recovered. A TEM image of the tri-layer sample is also shown in Fig. 4(c). The van-der-Waals planes of Bi$_2$Se$_3$ can be clearly resolved in this image along with the cubic crystal structure of EuSe, oriented mainly along the (001) direction. However, despite this dominant orientation, some (111) inclusions are observed.

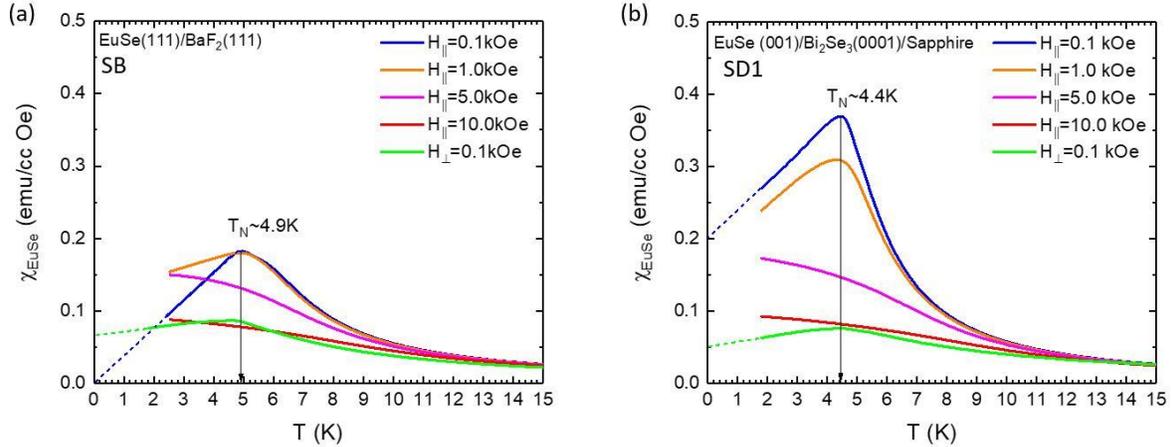

Fig. 5. Temperature dependence of magnetic susceptibilities of EuSe with in-plane $H_\parallel$ and out-of-plane $H_\perp$ magnetic fields for (a) EuSe (111) and (b) EuSe (001): Both samples show evolution of magnetic properties from AFM to FM when increasing applied magnetic field. The dashed lines indicate the comparison of zero temperature susceptibilities at small $H_\parallel$ and $H_\perp$ for both samples.

Overall, XRD measurements clearly evidence a (111) oriented growth on nearly lattice-match BaF$_2$ (111) and Pb$_{1-x}$Eu$_x$Se (111), but either a mixed phase or a (001)-oriented growth on GaAs(111)B and the Bi$_2$Se$_3$ surface with large lattice mismatch. This crystallographic orientation flexibility may provide and extra control of magnetic exchange coupling at a EuSe/Bi$_2$Se$_3$ interface and an extra knob to alter properties of TSS.

## Magnetic properties

Magnetic characterization of EuSe films is performed in a Quantum Design MPMS-3 SQUID magnetometer. Details of sample preparation and data analysis can be found in the Supplementary Material. This section is focused on the comparison between SB and SD samples grown on BaF$_2$ and Bi$_2$Se$_3$. Temperature dependence of magnetic susceptibility $\chi$ for SB and SD1 is plotted in Fig. 5(a) and 5(b) respectively. The Néel temperature $T_N = 4.4 - 4.9$ K of strain-free EuSe thin films SB and SD1 is found to be close to the $T_N$ in bulk crystals (4.6K) and in thick epitaxial films ($4.75 \pm 0.25\ K$) [17]. For EuSe (111), we also find that the in-plane $\chi_\parallel \to 0$ for $T \to 0$, while out-of-plane $\chi_\perp$ remains finite (Fig. 5(a)). This confirms that magnetization at low fields is lying within the (111) growth plane, perpendicular to the growth axis. For (001) growth of EuSe, both $\chi_\parallel$ and $\chi_\perp$ are finite at small magnetic fields when $T \to 0$, as shown by the extended dashed lines (Fig. 5(b)). This is consistent with magnetic field pointing at an angle to the magnetization axis for both field directions and indicates that magnetization does not lie along the growth axis but remains within the (111) plane.

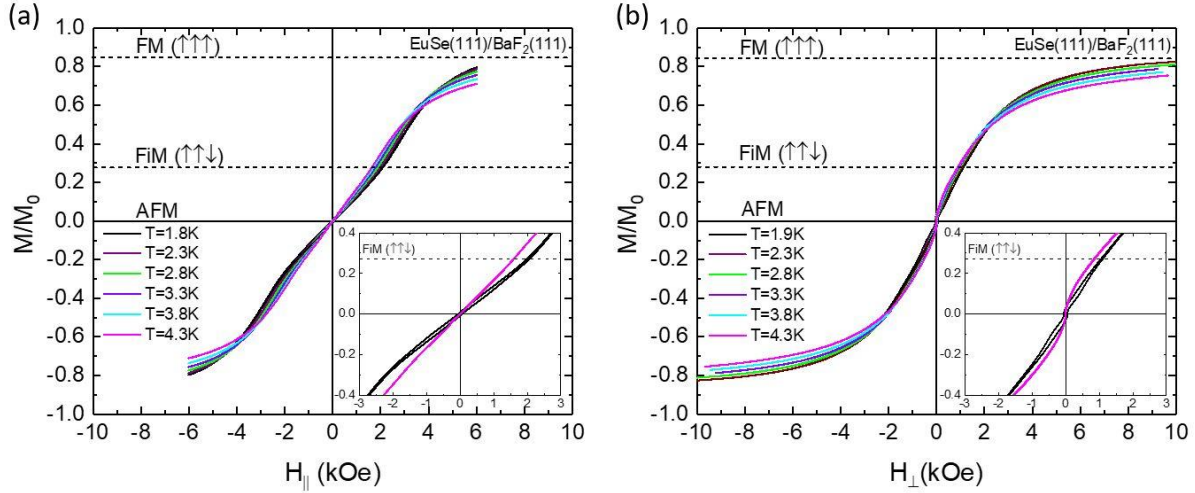

Fig. 6. Normalized magnetic moment per $Eu^{2+}$ ion as a function of magnetic field at different temperatures for (a) $H_\parallel$ and (b) $H_\perp$ for EuSe (111) on nearly-lattice-match $BaF_2$ (111). The ferromagnetic FM and ferrimagnetic FiM saturation values are indicated by dashed lines. Inset: hysteresis shown on an expanded scale.

In Fig. 6, we plot magnetization loops (M(H)) normalized by fully magnetized $Eu^{2+}$ (6.94$\mu_B$) for sample SB. Magnetization loops are shown for different temperatures and for field applied parallel ($H_\parallel$) and perpendicular ($H_\perp$) to (111) planes. Magnetic transitions are more pronounced in a plot of a derivative (dM/dH), as shown in Fig. 7. The four peaks correspond to the expected FM-FiM (A) and FiM-AFM (B) phase transitions. In agreement with previous reports FM-FiM transition is observed at $H_\parallel(A) = 2 - 3$ kOe and is not hysteretic. The sharpness of this transition does not depend on temperature, as can be seen from almost $T$-independent height of A peaks. Unexpectedly, AFM-FiM transition (peaks B) is found to be strongly hysteretic, the hysteresis is emphasized in Fig. 7a where a sweep up (a dashed line) is overlapped on down sweep curves. Temperature dependence of peaks B indicates that the AFM-FiM transition broadens as $T$ decreases. For $H_\perp$, differential magnetic susceptibility shows a large peak centered at $H_\perp = 0$, consistent with the canting of spins out-of-plane, and the observation of A and B peaks which correspond to FM-FiM and AFM-FiM transitions is influenced by small misalignment of the sample with respect to the field direction. Magnetic susceptibility of EuSe (001) films is a combination of $H_\parallel$ and $H_\perp$ response for the EuSe (111) film (see Supplemental Materials).

Using field dependence of A and B peaks positions we construct a magnetic phase diagram of thin EuSe (111) films in Fig. 7. For comparison, purple lines show phase boundaries in bulk EuSe

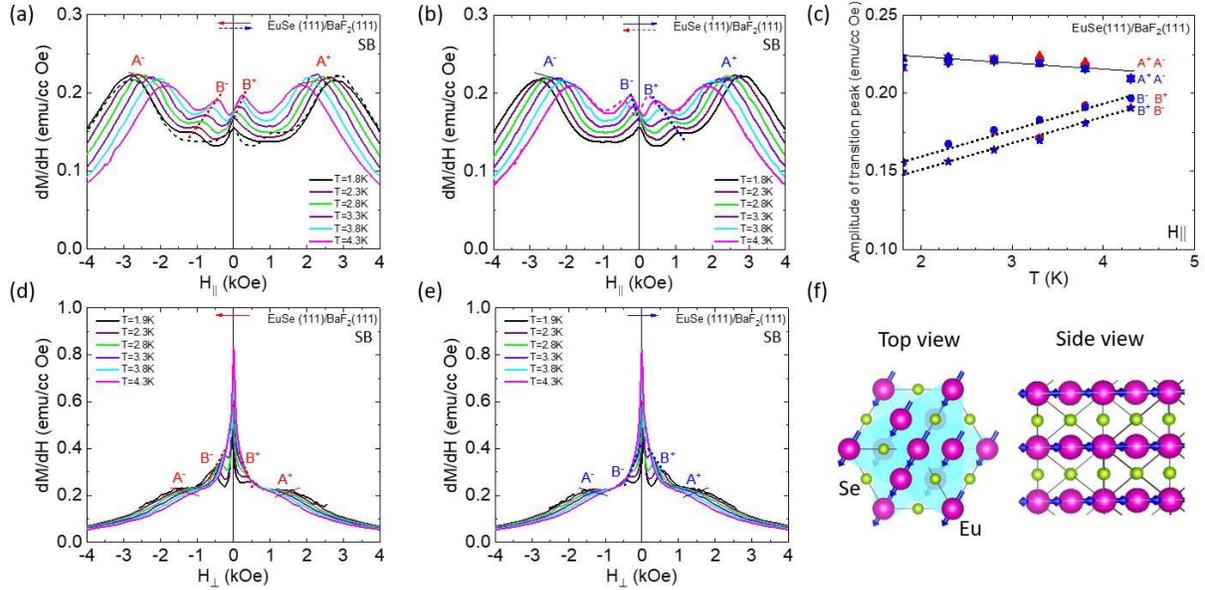

Fig. 7. Magnetic properties of EuSe (111): Differential magnetic susceptibilities (dM/dH) as a function of magnetic field with two opposite field sweep directions at different temperatures for (a) (b) in-plane $H_\parallel$ and (d) (e) out-of-plane $H_\perp$: FM-FiM transitions are labeled by $A^+$ and $A^-$ and guided by solid lines, AFM-FiM transitions are labeled by $B^+$ and $B^-$ and guided by dotted lines, arrows indicate the field sweep direction corresponding to the dM/dH curves; (c) Amplitude of transitions as a function of temperatures for in-plane $H_\parallel$; (f) EuSe (111) crystal structure with magnetization plane (cyan color) and spin rotation (blue arrows).

(111) extracted from Ref. [17]. We observe $\approx 20\%$ enhancement of field values for the FM-FiM. In bulk samples two distinct AFM regions with different magnetic periodicity are observed above 3.6 K and below 2.8K with no AFM phase formed in the temperature range 2.8-3.6 K. In thin films we found that AFM-FiM transition is present at all $T < T_N$, the transition field $H_\parallel(B^-)$ for field sweeping down uniformly *increases* as $T$ decreases. The value of $H_\parallel(B^-) \approx 0.4\, H_\parallel(A)$ of the FM-FiM transition and has similar *T*-dependence. In contrast, $H_\parallel(B^+)$ for the FiM-AFM transition *decreases* as $T$ decreases, $H_\parallel(B^+) \approx 0$ for $T < 3.2$ K. Small hysteresis in the band gap energy as a function of magnetic field has been reported in $\mu$m-thick films [25], large hysteresis in the current work ($\approx 1$ kOe) can be attributed to the strongly reduced film thickness compared to the previous work. Strong dependence of phase transition dynamics on a film thickness and the presence of surfaces has been extensively studied for the AFM-FM transition that occurs near room temperature in FeRh alloys [26-28]. Such strong modification of the AFM-FiM transition in thin EuSe films can also be reflected in the strength of EuSe/TI magnetic exchange.

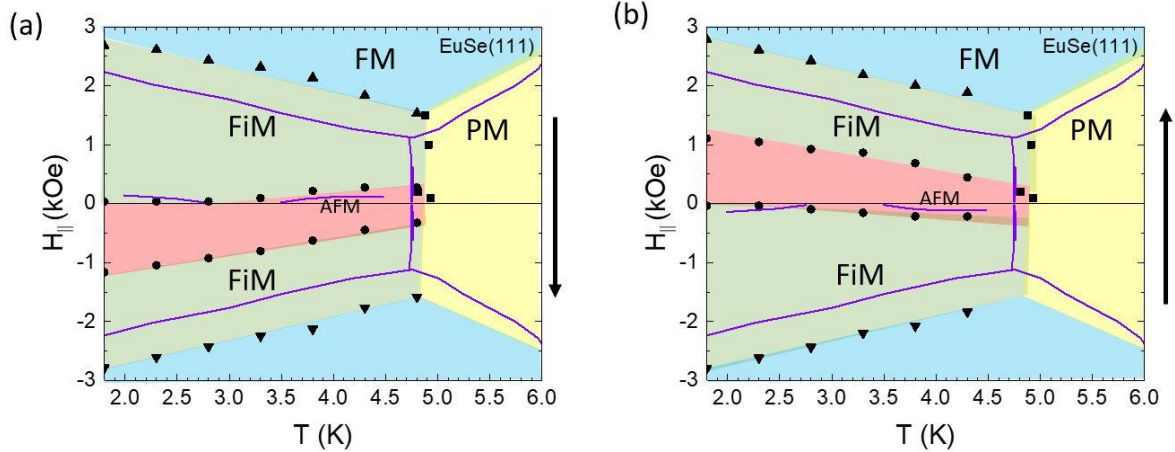

Fig. 8. In-plane phase diagram of EuSe (111) on nearly-lattice-match BaF$_2$ (111) for (a) field sweeping down and (b) field sweeping up. Dots and triangles are positions of A and B peaks in Figure 7. The red, green, blue and yellow areas indicate the AFM, FiM, FM and PM regions. Arrows indicate the field sweep directions. The purple lines display phase diagram of bulk (um thick) EuSe (111) reprinted from ref [17].

## Conclusion:

In summary, we have studied structural and magnetic properties of EuSe thin epilayers grown on different substrates and buffer layers. On BaF$_2$ (111), EuSe growth in (111) direction in a quasi-2D growth mode which results in rough surface. Introduction of Pb$_{1-x}$Eu$_x$Se buffer layers improves surface morphology. However, Pb$_{1-x}$Eu$_x$Se with $x < 0.1$ is conducting, which reduces its utility for TI exchange biasing. EuSe growth in (001) direction on both GaAs(111)B and Bi$_2$Se$_3$ substrates with good morphology. The magnetic phase diagram for thin films has AFM, FiM and FM phases similar to bulk crystals. Field value for the FiM-FM transitions is slightly enhanced compared to the bulk, while AFM-FiM transition is highly hysteretic and has *T*-dependence different from bulk crystals. Control of phase transition dynamics is a key for the successful realization of selective gaping of TSS in magnetic material/TI heterostructures. An observed enhancement of AFM-FiM transition and strong hysteresis might provide an applicable mean to control magnetic order and phase transition by tuning the film thickness. Also, (001) growth of EuSe grown on Bi$_2$Se$_3$ or GaAs (111) surfaces provides an opportunity to realize an AFM interface with TSS. To realize FM interface with TSS in EuSe-based topological heterostructures it is necessary to develop suitable underlayer TI layers with similar lattice structure and matched lattice parameters.

## Acknowledgments:

Authors acknowledge support by the NSF DMR-2005092 (Y.W. and L.P.R.); and by the NSF DMR-1905277 (X.L., S.K.B., J.K.F., B.A.A.) grants. M.Z. and T.O. acknowledge use of facilities for High

Resolution Electron Microscopy at University of Notre Dame. The data that support the findings of this study are available from the corresponding author upon reasonable request.

* these authors contributed equally.

**References:**

1. Hasan, M. Z. & Kane, C. L. Topological insulators. *Rev. Mod. Phys.* **82**, 3045–3067 (2010).

2. Yu, R. *et al.* Quantized Anomalous Hall Effect in Magnetic Topological Insulators. *Science.* **329**, 61–64 (2010).

3. Mogi, M. *et al.* Magnetic modulation doping in topological insulators toward higher-temperature quantum anomalous Hall effect. *Appl. Phys. Lett.* **107**, 182401 (2015).

4. Xiao, D. *et al.* Realization of the Axion Insulator State in Quantum Anomalous Hall Sandwich Heterostructures. *Phys. Rev. Lett.* **120**, 56801 (2018).

5. Fu, L., & Kane, C. L. Superconducting Proximity Effect and Majorana Fermions at the Surface of a Topological Insulator. *Phys. Rev. Lett.* **100**, 096407 (2008).

6. Xu, Y. *et al.* Observation of topological surface state quantum Hall effect in an intrinsic three-dimensional topological insulator. *Nat. Phys.* **10**, 956–963 (2014).

7. Rienks, E. D. L. *et al.* Large magnetic gap at the Dirac point in $Bi_2Te_3$/$MnBi_2Te_4$ heterostructures. *Nature* **576**, 423–428 (2019).

8. Deng, Y. *et al.* Quantum anomalous Hall effect in intrinsic magnetic topological insulator $MnBi_2Te_4$. *Science.* **367**, 895–900 (2020).

9. Chen, P. *et al.* Tailoring the Hybrid Anomalous Hall Response in Engineered Magnetic Topological Insulator Heterostructures. *Nano Lett.* **20**, 1731–1737 (2020).

10. Chang, S. J. *et al.* Heterostructured ferromagnet-topological insulator with dual-phase magnetic properties. *RSC Adv.* **8**, 7785–7791 (2018).

11. Wei, P. *et al.* Exchange-Coupling-Induced Symmetry Breaking in Topological Insulators. *Phys. Rev. Lett.* **110**, 186807 (2013).

12. Katmis, F. *et al.* A high-temperature ferromagnetic topological insulating phase by proximity coupling. *Nature* **533**, 513–516 (2016).

13. Figueroa, A. I. *et al.* Absence of Magnetic Proximity Effect at the Interface of $Bi_2Se_3$ and $(Bi,Sb)_2Te_3$ with EuS. *Phys. Rev. Lett.* **125**, 226801 (2020).

14. Wang, F. *et al.* Observation of Interfacial Antiferromagnetic Coupling between Magnetic Topological Insulator and Antiferromagnetic Insulator. *Nano Lett.* **19**, 2945–2952 (2019).

15. He, Q. L. *et al.* Tailoring exchange couplings in magnetic topological-insulator/antiferromagnet heterostructures. *Nat. Mater.* **16**, 94–100 (2017).

16. Mauger, A. & Godart, C. The magnetic, optical, and transport properties of representatives of a class of magnetic semiconductors: The europium chalcogenides. *Phys. Rep.* **141**, 51–176 (1986).

17. Lechner, R. T. *et al.* Strain induced changes in the magnetic phase diagram of metamagnetic heteroepitaxial EuSe/PbSe$_{1-x}$Te$_x$ multilayers. *Phys. Rev. Lett.* **94**, 157201 (2005).

18. Lechner, R. T. *et al.* Spin configurations in strained magnetic EuSe/PbSe$_{1-x}$Te$_x$ superlattices grown by molecular beam epitaxy. *Phys. E Low-Dimensional Syst. Nanostructures* **32**, 379–382 (2006).

19. Griessen, R., Landolt, M. & Ott, H. R. A new antiferromagnetic phase in EuSe below 1.8 K. *Solid State Commun.* **9**, 2219–2223 (1971).

20. Köbler, U. *et al.* Biquadratic exchange interactions in the Europium monochalcogenides. *Zeitschrift fur Phys. B-Condensed Matter* **100**, 497–506 (1996).

21. X. Liu, J. Wang, L. Riney, S.K. Bac, D.J. Smith, M.R. McCartney, I. Khan, A.J. Hoffman, M. Dobrowolska, J. Furdyna, B. A. A. Unraveling the structural and electronic properties of the strained PbSe on GaAs. *Accepted Journal of Crystal Growth* (2021).

22. Richardella, A. *et al.* Coherent heteroepitaxy of Bi2 Se3 on GaAs (111)B. *Appl. Phys. Lett.* **97**, 262104 (2010).

23. Liu, X. *et al.* Structural properties of Bi$_2$Te$_3$ and Bi$_2$Se$_3$ topological insulators grown by molecular beam epitaxy on GaAs(001) substrates. *Appl. Phys. Lett.* **99**, 171903 (2011).

24. Vishwanath, S. *et al.* Controllable growth of layered selenide and telluride heterostructures and superlattices using molecular beam epitaxy. *J. Mater. Res.* **31**, 900–910 (2016).

25. You, A., Be, M. A. Y. & In, I. Hysteresis loops of the energy band gap and effective factor up to 18 000 for metamagnetic epilayers. *Appl. Phys. Lett.* **85**, 67 (2004).

26. Suzuki, I., Koike, T., Itoh, M., Taniyama, T. & Sato, T. Stability of ferromagnetic state of epitaxially grown ordered FeRh thin films. *J. Appl. Phys.* **105**, 103–106 (2009).

27. Yu, C. Q. *et al.* Thickness-dependent magnetic order and phase-transition dynamics in epitaxial Fe-rich FeRh thin films. *Phys. Lett. A* **383**, 2424–2428 (2019).

# Epitaxial growth and magnetic characterization of EuSe thin film with various crystalline orientations

Ying Wang[1*], Xinyu Liu,[2*] Seul-Ki Bac,[2] Jacek K. Furdyna,[2] Badih A. Assaf,[2] Maksym Zhukovskyi,[3] Tatyana Orlova,[3] Neil R Dilley[4], Leonid P. Rokhinson[1,4,5]

**Supplementary Material**

## Atomic force microscopy (AFM) images of sample surface

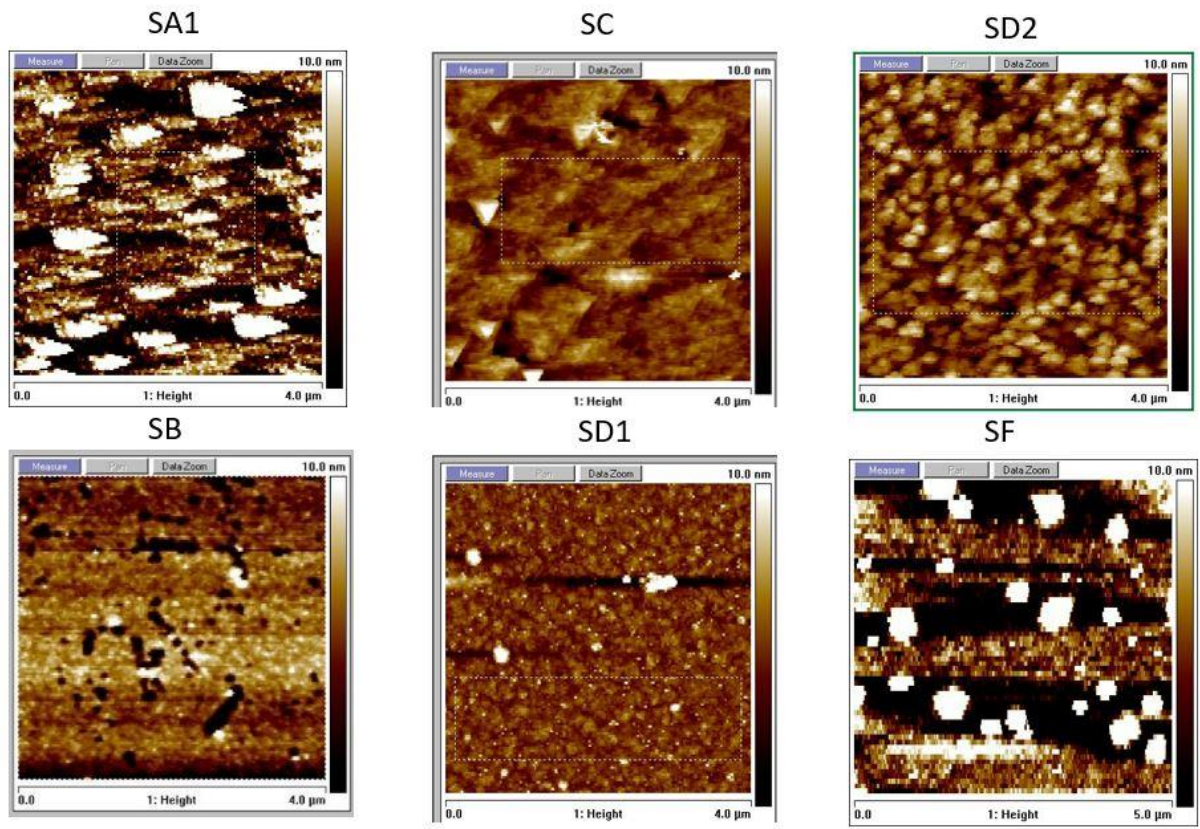

FIG S1. AFM images of samples listed in the Table 1 of the main text.

## Structural properties of EuSe grown on GaAs(111)B

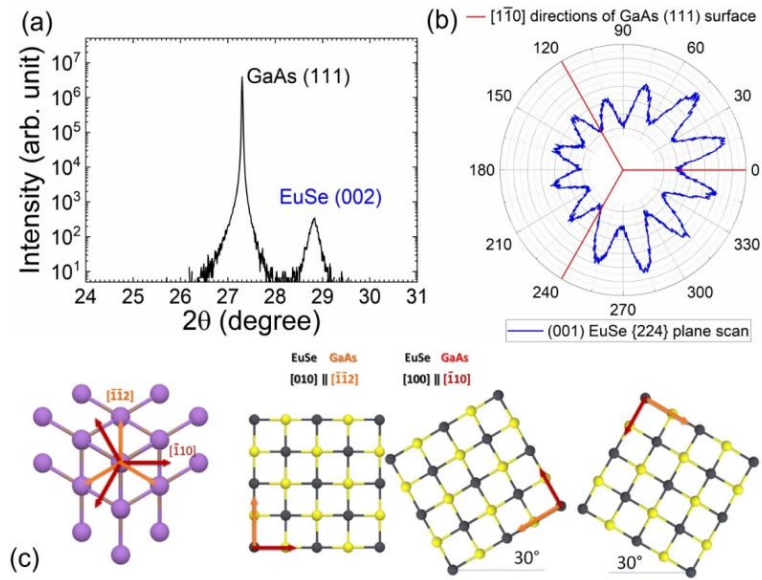

FIG S2. (a) High-resolution XRD ω-2θ scan of the (111) Bragg peak of GaAs substrate and (002) Bragg peak of EuSe. (b) $\varphi$-scan of the (224) EuSe Bragg peak. (c) $[010] \parallel [\bar{1}\bar{1}2]$ ($[100] \parallel [\bar{1}10]$) rocksalt-zincblende in-plane alignment.

FIG S2 (a) shows an XRD ω-2θ scan obtained for Sample SA1 grown on a GaAs (111)B substrate. A strong (002) EuSe peak is observed, indicating that a (001)-oriented EuSe layer is indeed obtained despite the (111) GaAs surface. A $\varphi$-scan is also performed about a (224) EuSe peak (FIG S2 (b)), a 12-fold symmetric reflection pattern is visible about the EuSe peak. FIG S2 (c) presents the alignment of EuSe(001) on the GaAs(111) surface as follows: If the [010] direction of EuSe is aligned with the $[\bar{1}\bar{1}2]$ direction of GaAs, then a [100] direction of EuSe is aligned with the $[\bar{1}10]$ direction of GaAs. There are, however, three ⟨112⟩ directions yielding 3 possible alignments shown in the FIG S2 (c). For each one of these cases there are 4 possible alignments of the rock-salt unit cell yielding 12 possible orientations.

FIG S3 (a) shows an XRD ω-2θ scan obtained for Sample SA2 also grown on a GaAs (111)B substrate. Both (111) and (002) EuSe peaks are observed, indicating that a mixed (111)/(001)-oriented EuSe layer is formed in this case. A $\varphi$-scan is performed about a (224) EuSe peak in respect to EuSe with (111) orientation (FIG S3 (b)), a 3-fold symmetric reflection pattern is visible about such EuSe peak, suggesting $[\bar{1}10] \parallel [\bar{1}10]$ rocksalt-zincblende in-plane alignment for EuSe film with (111) growth direction (FIG S3 (d)). A $\varphi$-scan is also performed about a (224) EuSe peak respect to EuSe with (001) orientation (FIG S3 (c)), a 12-fold symmetric reflection pattern is observed, including 6-fold strong pattern and 6-fold weak pattern. In this case, FIG S3 (e) presents the alignment of EuSe(001) inclusion on the GaAs(111) surface as follows: If the $[\bar{1}10]$ direction of EuSe is aligned with the $[\bar{1}\bar{1}2]$ direction of GaAs, then a [110] direction of EuSe is aligned with the $[\bar{1}10]$ direction of GaAs. Three ⟨112⟩ directions yield 3 possible alignments shown in the FIG S3 (e). For each one of these cases there are 4 possible alignments of the rock-salt unit cell yielding 12 possibilities. Due to the effect from EuSe grown on (111) direction, which usually presents 6-fold <110> planes, the 12-fold symmetric reflection pattern evolves into a combination of 6-fold strong pattern and 6-fold weak pattern.

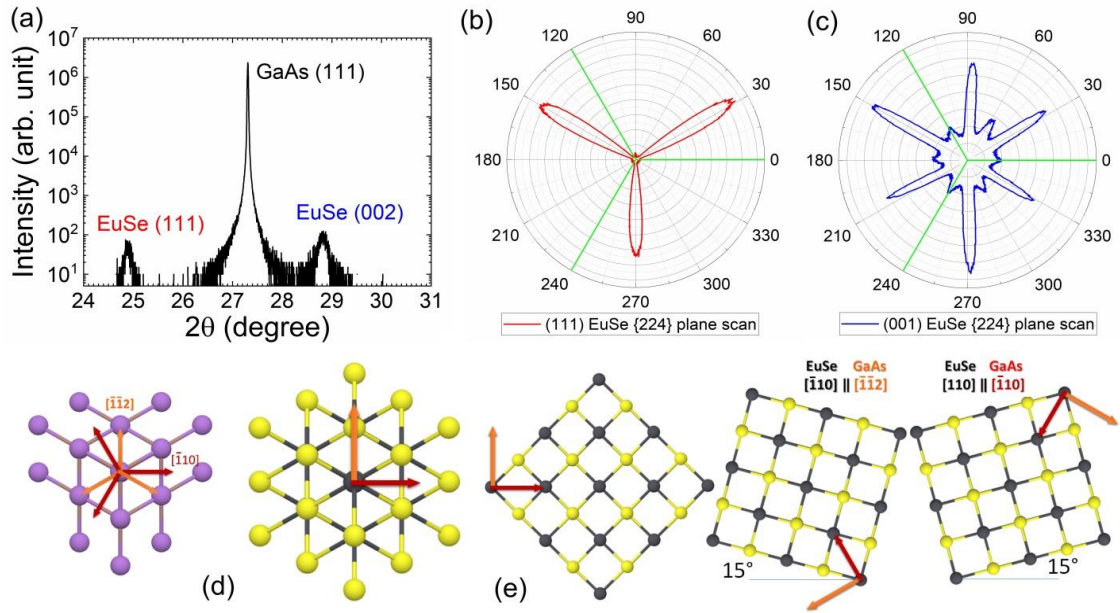

FIG S3. (a) High-resolution XRD ω-2θ scan of the (111) Bragg peak of GaAs substrate, and (111) and (002) Bragg peaks of EuSe. (b) $\varphi$-scan of the (224) EuSe Bragg peak with (111) growth direction. (c) $\varphi$-scan of the (224) EuSe Bragg peak with (001) growth direction. (d) $[\bar{1}10] \parallel [\bar{1}10]$ rocksalt-zincblende in-plane alignment for EuSe film with (111) growth direction. (e) $[\bar{1}10] \parallel [\bar{1}\bar{1}2]$ ($[110] \parallel [\bar{1}10]$) rocksalt-zincblende in-plane alignment for EuSe inclusion with (001) growth direction.

## Magnetic measurements

The magnetic characterization of EuSe films were performed in a Quantum Design MPMS-3 SQUID magnetometer. All the samples were carefully cleaned with isopropanol and we used standard plastic translucent drinking straws as wafer holders. Samples were mounted in the straws such that the magnetic field was applied either in-plane ($H_{\parallel}$) or out-of-plane ($H_{\perp}$). The magnetization loops (HLs) were obtained at various fixed temperatures below and above Néel temperature. All magnetization measurements were corrected to account for the demagnetizing field of thin film (demagnetization factors $N_{\parallel} = 0$ and $N_{\perp} = 1$). In addition, a correction to the reported applied magnetic field was made to HLs measurement data and arises due to remnant magnetic field in the superconducting magnet (See Quantum Design Application Note 1500-011). The field correction was measured and verified using a clean paramagnetic sample (a single crystal of gadolinium gallium garnet) which is known to have a linear and reversible magnetization curve. For the sweep rates used in this studies the real fields is found to be lagging by 25 Oe, and appropriate correction is applied to all the reported data.

## An example of normalization of magnetic moment per $Eu^{2+}$ ion

1) **Subtraction of substrate magnetization**
   The total magnetizations ($m_{total}$) of sample SD1 were displayed in FIG S4 (a)(d). The magnetization of a substrate should be subtracted to obtain magnetization of the thin film. Magnetizations of the sapphire ($m_{sapphire}$) substrate is measured after removing the top EuSe epilayers using a plastic blade, FIG S4 (b)(e). Magnetizations of EuSe is obtained as $m_{EuSe} = m_{total} - m_{sapphire}$.

2) **Normalization of magnetic moment per $Eu^{2+}$ ion**
   Magnetization of EuSe can be converted to $\frac{\mu_B}{Eu}$ by using the following relationship:
   $$\frac{emu}{cc} = \frac{1.0783 \cdot 10^{20} \frac{\mu_B}{4}}{(\lambda_{EuSe} \cdot 10^{-7})^3 Eu} = 6.41 \cdot 10^{-3} \frac{\mu_B}{Eu},$$ where the lattice spacing $\lambda_{EuSe} = 0.6186$ nm is measured by XRD. Completely polarized $Eu^{2+}$ ion in EuSe is expected to have $M_0 = 6.94 \frac{\mu_B}{Eu}$ (from DFT studies[1]).

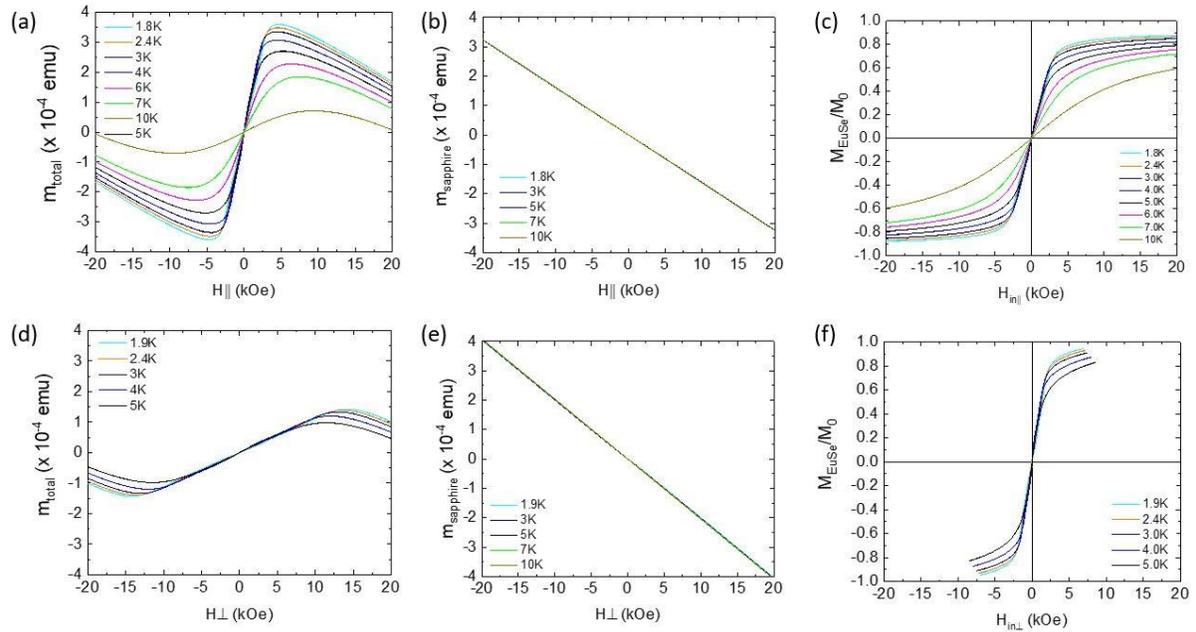

FIG S4. Total magnetic moment of (a)(d) EuSe on 2-dimentional van der Waals $Bi_2Se_3$ epilayers grown on sapphire (c-plane) substrate and (b)(e) pure sapphire substrate for (a)(b) $H_\parallel$ and (d)(e) $H_\perp$; Normalized magnetic moment per $Eu^{2+}$ ion as a function of magnetic field after correction of demagnetization field at different temperatures for (c) $H_{in\parallel}$ and (f) $H_{in\perp}$.

# Phase diagram for EuSe (001) (EuSe/ Bi$_2$Se$_3$ /sapphire)

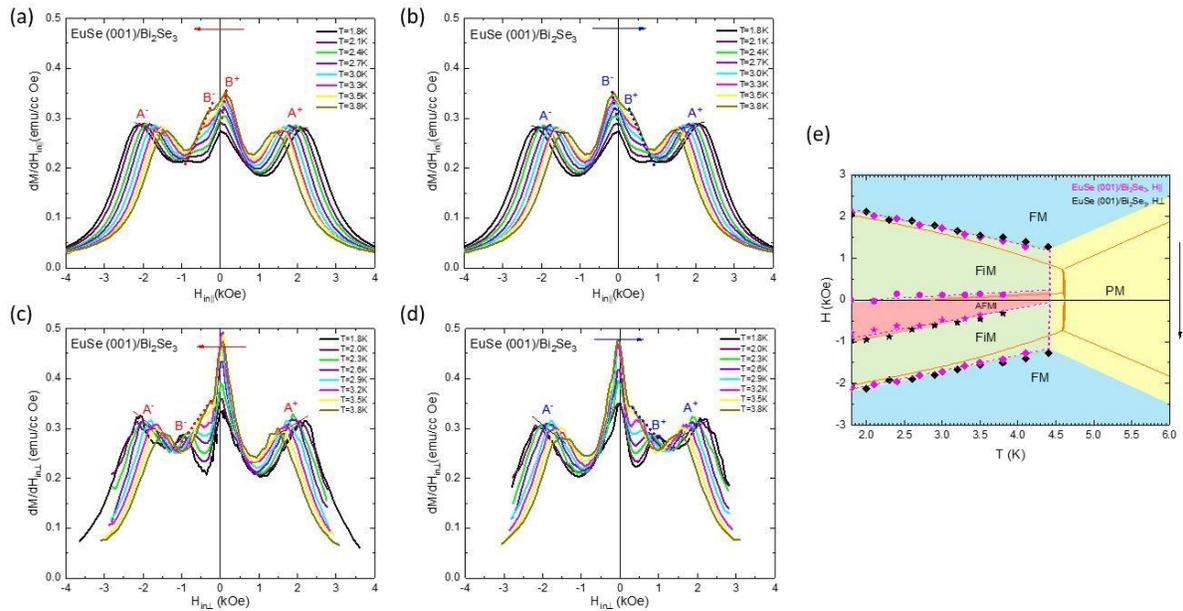

FIG S5. In-plane (a)(b) and out-of-plane (c)(d) differential magnetic susceptibility dM/dH of EuSe Bi$_2$Se$_3$ epilayers grown on sapphire (c-plane) substrate as a function of magnetic field . FM-FiM transitions are labeled by $A^+$ and $A^-$, AFM-FiM transitions are labeled by $B^+$ and $B^-$; (e) Corresponding phase diagram for one field sweep direction: magenta and black symbols indicate $H_\parallel$ and $H_\perp$, orange curves display phase diagram for a single crystal EuSe reported in ref [2].

Magnetic phase transitions for in- and out-of-plane magnetic field in sample SD1 are extracted from the peak positions in differential magnetic susceptibility dM/dH, FIG S5. Similarities between $dM/dH$ data for $H_\parallel$ and $H_\perp$ field directions, and the presence of a peak around $H = 0$ confirms that in thin (001) EuSe films magnetization is retained within the (111) planes.

**References:**

1. J. Kim, K.-W. Kim, H. Wang, J. Sinova, and R. Wu, Understanding the Giant Enhancement of Exchange Interaction in Bi$_2$Se$_3$/EuS Heterostructures, *Phys. Rev. Lett.* **119**, 027201 (2017).
2. Griessen R, Landolt M, Ott HR. A new antiferromagnetic phase in EuSe below 1.8 K. *Solid State Commun*. **9**, 2219–2223 (1971).